\documentclass[conference,a4paper]{IEEEtran}
 
\IEEEoverridecommandlockouts
\usepackage{cite}
\usepackage{amsmath,amssymb,amsfonts}
\usepackage{algorithmic}
\usepackage{graphicx}
\usepackage{textcomp}
\usepackage{xcolor}
\def\BibTeX{{\rm B\kern-.05em{\sc i\kern-.025em b}\kern-.08em
    T\kern-.1667em\lower.7ex\hbox{E}\kern-.125emX}}

    \usepackage{tikz}
\tikzstyle{pinstyle} = [pin edge={to-,thin,black}]
\usetikzlibrary{calc}
\newlength\fheight
\newlength\fwidth
\usepackage{pgfplots}
\usepackage{array}
\newcolumntype{?}{!{\vrule width 1pt}}
\usepackage{booktabs}
\usepackage{stfloats}
\usepackage{multicol}
\usepackage{arydshln}
\pgfplotsset{compat=newest}
\usepackage[hyphens]{url}
\usepackage{hyperref}
\usepackage[hyphenbreaks]{breakurl}

\usepackage{textcomp}
\usepackage[absolute]{textpos}

\newcommand{\copyrightstatement}{
    \begin{textblock}{15}(0.5,0.3)    
         \noindent
         \centering
         \textblockcolour{white}
         \footnotesize
         \copyright 2020 IEEE. Personal use of this material is permitted. Permission from IEEE must be obtained for all other uses, in any current or future media, including reprinting/republishing this material for advertising or promotional purposes, creating new collective works, for resale or redistribution to servers or lists, or reuse of any copyrighted component of this work in other works
    \end{textblock}
}

\begin{document}

\copyrightstatement
\title{A Comparative Analysis of the Time and Energy Demand of Versatile Video Coding and High Efficiency Video Coding Reference Decoders}

\author{\IEEEauthorblockN{Matthias Kr\"anzler, Christian Herglotz, and Andr\'e Kaup}
\IEEEauthorblockA{\textit{Multimedia Communications and Signal Processing}\\
		\textit{Friedrich-Alexander University Erlangen-N\"urnberg (FAU)}\\
Erlangen, Germany\\
\{matthias.kraenzler, christian.herglotz, andre.kaup\}@fau.de}
}

\maketitle
\renewcommand{\arraystretch}{1.2}

\begin{abstract}

This paper investigates the decoding energy and decoding time demand of \mbox{VTM-7.0} in relation to HM-16.20. We present the first detailed comparison of two video codecs in terms of software decoder energy consumption. The evaluation shows that the energy demand of the VTM decoder is increased significantly compared to HM and that the increase depends on the coding configuration. For the coding configuration randomaccess, we find that the decoding energy is increased by over 80$\%$ at a decoding time increase of over 70$\%$. Furthermore, results indicate that the energy demand increases by up to 207$\%$ when Single Instruction Multiple Data (SIMD) instructions are disabled, which corresponds to the HM implementation style. By measurements, it is revealed that the coding tools MIP, AMVR, TPM, LFNST, and MTS increase the energy efficiency of the decoder. Furthermore, we propose a new coding configuration based on our analysis, which reduces the energy demand of the VTM decoder by over 17\% on average.

\end{abstract}

\IEEEpubid{\begin{minipage}[t]{\textwidth}\ \\[10pt] 978-1-7281-9320-5/20/\$31.00~\copyright~2020 IEEE\end{minipage}}

\section{Introduction}
\label{sec:intro}

The viewing of online videos has a significant influence on the global CO2 emissions. In 2018, the viewing of online video content had a share of $1\%$ of the global greenhouse gas emissions, which is comparable to the emissions of Spain~\cite{Efoui-Hess2019}. As the amount of data that is transmitted over the internet will increase every year by $26\%$~\cite{CSI2019}, it can be expected that the influence of online video content will increase within the next years, too. Energy efficiency is not only environmentally essential but also practically important on mobile devices like smartphones, which are powered by a battery.

In 2020, the first version of the next-generation video standard Versatile Video Coding (VVC) will be standardized to reduce the bit rate by $50\%$~\cite{Sullivan2017} in relation to the state-of-the-art video codec High Efficiency Video Coding (HEVC). However, the improvement of the rate-distortion (RD) efficiency is at the cost of higher computational complexity and a higher energy demand.

In literature, the complexity of HEVC was compared to the predecessor video codec Advanced Video Coding (AVC). In \cite{Viitanen2012}, the authors compared the computational complexity of both reference software implementation of HEVC and AVC in terms of cycles counts per frame. It was determined that the complexity of HEVC is increased by an absolute value of over 80\% for the RA configuration. In \cite{Bossen2012}, the reference software encoder and decoder of HEVC, which is the HM software, was analyzed in detail. Therefore, the authors determined the time consumption of several functions of the decoder. For the randomaccess configuration, the main contributors of the decoding time are interpolation and loop filter, which contribute more than one third of the total time.

Currently, there is no comparative analysis of different video codecs in terms of the energy demand of the decoder, which we perform in this paper. Therefore, we consider the distortion of the bit streams in terms of PSNR for the comparison of the energy by using Bj{\o}ntegaard-Delta metrics. 

\begin{figure}[!t]
	\label{fig:BlockChart}
	\centering
	\setlength\fheight{4cm}
	\setlength\fwidth{7.5cm}
	\begin{tikzpicture}
	\begin{scope}[minimum width=15mm,minimum height=9mm,align=center]
	\node[draw,text width=1.9cm] (v1) at (0,0) {Encoder};
	\node[draw,text width=1.9cm] (v2) at (3,0) {Decoder};
	\node[draw,text width=1.9cm] (v4) at (0,-2) {Optimization};
	\node[draw,text width=1.9cm] (v3) at (3,-2) {Analysis};	
	\node[text width=3cm] (v5) at ($(v2) + (0,1)$) {Power meter};
	\draw[thick,rounded corners=8pt] (1.5,0) -- (1.5,1.5) -- (4.5,1.5) -- (4.5,-1) -- (1.5,-1) -- (1.5,0) ;
	\node[inner sep=0pt] (image1) at (-3,0){\includegraphics[width=2.6cm]{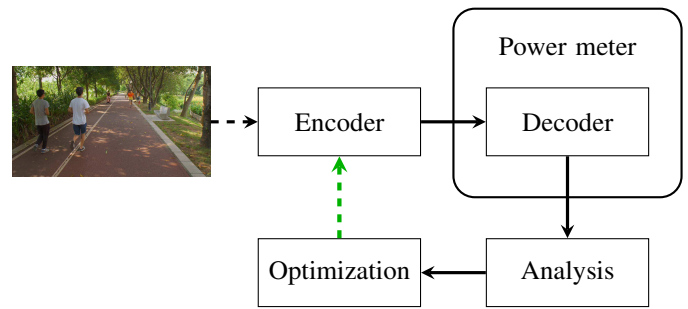}};
	
	\end{scope}
	\draw[-stealth,very thick] (v1) -- (v2);
	\draw[-stealth,very thick] (v2) -- (v3);
	\draw[-stealth,very thick] (v3) -- (v4);
	\draw[-stealth,dashed,ultra thick,black!30!green] (v4) -- (v1);
	\draw[-stealth,dashed,very thick,] (image1) -- (v1);
	
	\end{tikzpicture}
	\caption{Overview of the evaluation procedure for coded video sequences in this paper. First, the sequences are coded by a video encoder. Then, the generated bit streams are decoded by the corresponding decoder. Simultaneously, the energy demand of this process is measured by a power meter. The results of the measurement are analyzed by comparing, e.g., two video codecs. Finally, the conclusions of the analysis are used to optimize the energy demand of the VVC decoder by changing the encoding configuration.}
\end{figure}

For the decoding energy optimization, there are different approaches in the literature. In a first approach \cite{MallikarachchiTalagalaH.EtAl2017}, the complexity of a decoder is measured in terms of the number of CPU cycles that are necessary to decode a bit stream. The resulting information is used to estimate the decoding complexity during encoding and subsequently used to reduce the complexity of the decoder. Thereby, the complexity is reported to be reduced by 40\%. In the approach of \cite{HerglotzHeindelKaup}, the Decoding-Energy Rate-Distortion Optimization (DERDO) algorithm is proposed in HEVC. This algorithm uses bit stream feature-based models to estimate the energy demand of the decoder during the encoding. The RD-optimization is extended by the estimated energy demand and optimizes on the three metrics rate, distortion, and decoding energy. The authors show that the energy demand can be reduced by up to 30\% if the bit rate is increased by the same order. For VVC and HEVC, different models estimate accurately the energy demand of the decoder, which is necessary for the optimization of the energy demand \cite{Kraenzler2019}, \cite{Kraenzler2020}. Although sophisticated models exist, a detailed comparison of the energy demand has not been done before to the best of our knowledge. 

In this paper, the additional computational complexity of VVC compared to HEVC is investigated in terms of the decoding energy and time demand. The evaluation toolchain is shown in Figure~\ref{fig:BlockChart}. As an encoder, we use the HM and VTM encoder to generate bit streams. Then, the decoding process of the bit streams is measured in terms of energy and time demand. Afterwards, the measurements are used to analyze both video codecs in detail. Finally, the results of the analysis can be used to optimize the energy demand of the VVC decoder.  

From our in-depth analysis, we have the following findings. First, we determine that the decoding energy is increased by a higher degree than the decoding time. Second, we discover that the usage of SIMD instructions reduces the time demand more than the energy demand. Third, we show that several coding tools optimize the energy demand of the VVC decoder. Finally, we propose a new energy-efficient coding configuration that reduces the energy demand of the VTM decoder by 17\% on average.

This paper is organized as follows. First, Section~\ref{sec:2} explains the measurement setup and the used software setup. Afterwards, in Section~\ref{sec:3} we compare the decoding time and energy demand of VVC and HEVC. The influence of SIMD instructions on the decoding energy and time demand is evaluated for VVC in Section~\ref{sec:4}, and the influence of some coding tools is shown in Section~\ref{sec:5}. The improvement of the decoding energy of VVC is presented in Section~\ref{sec:6}. Finally, Section~\ref{sec:7} concludes this paper.

\section{Energy measurement and test set}
\label{sec:2}

\begin{table*}[t!]
\caption{Evaluation of BDR, BDDE, and BDDT for three encoding configurations. The reference is the measurement with HM-16.20.}
\label{tab:HEVCVVC}
\begin{center}
\vspace{-.3cm}
\begin{tabular}{c ? c |  c  | c  ? c |  c  | c  ? c |  c  | c }
         & \multicolumn{3}{c?}{AI over HM-16.20}  & \multicolumn{3}{c?}{LB over HM-16.20}  & \multicolumn{3}{c}{RA over HM-16.20}\\
Class & BDR in \% &  BDDT in \% & BDDE in \% &  BDR in \% & BDDT in \% & BDDE in \% &  BDR in \% & BDDT in \% & BDDE in \%  \\
	   \hline \hline
 A1 & $-27.77$ & $51.11$ & $55.51$ & $-$ & $-$ & $-$ & $-37.65$ & $60.22$ & $70.31$ \\
A2 & $-24.19$ & $63.66$ & $67.10$ & $-$ & $-$ & $-$ & $-40.11$ & $72.08$ & $83.59$ \\
 B  & $-20.76$ & $64.31$ & $67.95$ & $-29.89$ & $52.66$ & $58.77$ & $-34.01$ & $59.96$ & $70.53$ \\
 C  & $-21.77$ & $70.87$ & $75.75$ & $-27.60$ & $71.53$ & $81.77$ &  $-28.89$ & $73.58$ & $84.79$ \\
 D  & $-17.40$ & $71.84$ & $81.33$ & $-24.56$ & $114.94$ & $134.35$ &  $-26.49$ & $96.17$ & $111.65$ \\
 E  & $-25.31$ & $50.88$ & $55.70$ &$-31.78$ & $51.12$ & $59.55$ & $-$ & $-$ & $-$ \\
 F  & $-38.76$ & $55.89$ & $60.62$ &$-41.68$ & $47.67$ & $54.58$ & $-41.09$ & $64.61$ & $74.71$ \\
\hline
Mean     & $-24.90$ & $62.03$ & $67.13$ & $-31.01$ & $67.66$ & $77.76$ &  $-34.31$ & $71.05$ & $82.56$ \\
\end{tabular}
\end{center}
\vspace{-0.5cm}
\end{table*}

The energy demand for the decoding of bit streams is determined by two measurements with Running Average Power Limit~(RAPL) according to the descriptions of~\cite{HerglotzSpringerReichenbachEtAl2018}. First, the energy demand during the decoding of the bit stream is measured. Afterwards, the energy demand during idle is measured and subtracted from the first measurement. For our measurements, we use a desktop PC with an \mbox{Intel~i7-8700} CPU, which is a hexa-core CPU with a base frequency of \mbox{3.20 GHz}. The measurements of RAPL can be accessed through the file system of an Unix-PC. By using RAPL, no external power meter is needed.

As the measurements are influenced by different disturbances, e.g., the ambient temperature or background processes of the operating system that are not related to the decoding process, we assume that it is not sufficient to obtain the true decoding energy $E_{\text{dec}}$ by measuring a single iteration. To ensure the validity of measurements, we perform multiple measurements for each bit stream and a confidence interval test on the mean energy as shown in \cite{HerglotzSpringerReichenbachEtAl2018}. For $\alpha$, we use a value of 0.99 and for $\beta$, a value of 0.02.

For the comparison of both video codecs, we use the reference software of HEVC, which is HM-16.20~\cite{HM1620} and VTM-7.0~\cite{VTM7} for VVC. For the evaluation, we follow the common test conditions for standard dynamic range~(SDR) video given in \cite{JVET-N1010}, which include 26 sequences. Furthermore, we distinguish between three different coding configurations, which are \mbox{All-Intra (AI)}, \mbox{Lowdelay B (LB)}, and \mbox{Randomaccess (RA)}. For each sequence and configuration, the bit streams are encoded with a Quantization Parameter (QP) of 22, 27, 32, and 37. Furthermore, we use an internal coding bit depth of 10~bit for VVC and HEVC.

\section{Comparison of VVC and HEVC}
\label{sec:3}

\begin{table*}[t!]
\caption{Results of the measurement of disabling SIMD optimizations for the decoding of bit streams. The reference is HM-16.20.}
\label{tab:SIMD}
\begin{center}
\vspace{-.3cm}
\begin{tabular}{c ? c | c ? c | c ? c | c}
         & \multicolumn{2}{c?}{AI over HM-16.20} & \multicolumn{2}{c?}{LB over HM-16.20} & \multicolumn{2}{c}{RA over HM-16.20}\\
 Class	     &  BDDT in \%  & BDDE in \%  &  BDDT in \%  & BDDE in \%  & BDDT in \%  & BDDE in \% \\
	     \hline \hline
 A1  & $136.28$ & $125.93$ & $-$      & $-$      & $226.70$  & $219.35$\\
 A2  & $133.31$ & $122.20$ & $-$      & $-$      & $258.60$  & $249.26$\\
 B   & $128.51$ & $118.07$ & $175.50$ & $166.47$ & $211.23$  & $211.50$\\
 C   & $116.45$ & $107.52$ & $174.73$ & $170.61$ & $207.72$  & $203.08$\\
 D   & $111.69$ & $105.92$ & $214.46$ & $215.51$ & $234.49$  & $229.00$\\
 E   & $121.68$ & $113.81$ & $116.10$ & $115.53$ & $-$       & $-$\\
 F   & $115.27$ & $107.04$ & $118.01$ & $113.48$ & $147.39$  & $142.58$\\
\hline
Mean & $122.69$ & $113.77$ & $162.73$ & $158.84$ & $211.76$  & $207.04$\\
\end{tabular}
\end{center}
\vspace{-0.6cm}
\end{table*}

In this section, we will compare both video codecs in terms of the RD performance, the time demand, and the energy demand of the decoder. Therefore, we use the YUV-PSNR, which is calculated according to~\cite{Ohm2012}, to evaluate the distortion.

We determine the Bj{\o}ntegaard-Delta rate (BDR) \cite{VCEG-M33}, for the comparison of the RD efficiency of HEVC and VVC. BDR indicates the mean bit rate savings in percent of a video codec in relation to another codec at equal objective visual quality. For the comparison of the decoding energy demand, we calculate the Bj{\o}ntegaard-Delta Decoding Energy (BDDE)~ \cite{HerglotzHeindelKaup}, which corresponds to the energy savings in percent at equal objective visual quality, similar to BDR. For the calculation of BDDE, we replace the bit rate by the decoding energy demand. Similarly, the Bj{\o}ntegaard-Delta Decoding Time (BDDT) evaluates the additional complexity in terms of the time demand. 

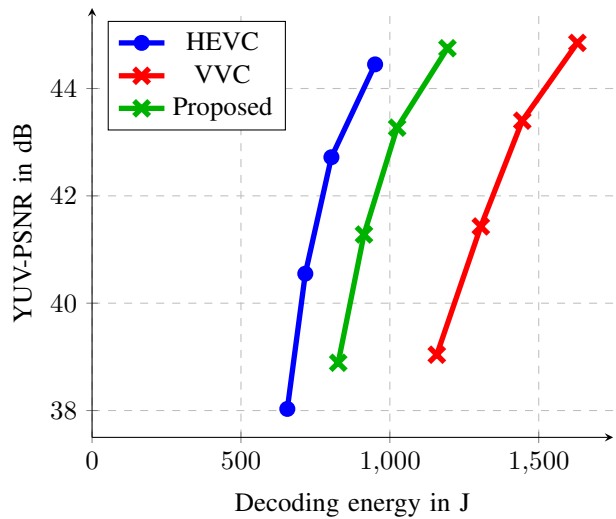
\begin{figure}[t!]
\begin{tikzpicture}
\begin{axis}[
    xlabel={Decoding energy in J},
    ylabel={YUV-PSNR in dB},
    xmin=0, xmax=1750,
    ymin=37.5, ymax=45.5,
    legend pos=north west,
    ymajorgrids=true,
    xmajorgrids=true,
    grid style=dashed,
    axis lines = left,
]

\addplot[
    color=blue,
    mark=*,
    line width=2pt,
    ]
    coordinates {
    (655.7,38.03)(716.1,40.55)(803.8,42.72)(950.5,44.45)
    };
    \addlegendentry{HEVC}
    
\addplot[
    color=red,
    mark=x,
    mark options={scale=2, fill=red},
    line width=2pt,
    ]
    coordinates {
    (1157.7,39.04)(1305.5,41.43)(1444.8,43.40)(1629.8,44.85)
    };
    \addlegendentry{VVC}

\addplot[
    color=black!30!green,
    mark=x,
    mark options={scale=2, fill=red},
    line width=2pt,
    ]
    coordinates {
    (826.60,38.89) (913.13,41.28) (1024.5,43.27) (1193.4,44.75)
    };
    \addlegendentry{Proposed}
 
\end{axis}
\end{tikzpicture}
\vspace*{-0.3cm}
\caption{Measurement of the decoding energy of the sequence FoodMarket4 with RA configuration with HM (blue) and VTM decoder (red). Our proposed encoding configuration for the VTM decoder is shown by the green line.}
\label{fig:BDDE}
\vspace{-0.4cm}
\end{figure}

In Figure~\ref{fig:BDDE}, the energy demand for the sequence FoodMarket4 with the reference RA configuration is shown vs. YUV-PSNR for the HM decoder (blue curve) and the VTM decoder (red curve). On the \mbox{x-axis}, the measured decoding energy demand is given in Joule, and on the \mbox{y-axis}, the YUV-PSNR is given in dB. The BDDE for these bit streams is $72.39\%$, which corresponds to an increase of $72.39\%$ in decoding energy demand for an equal objective visual quality. Our proposed encoding configuration, which will be discussed in detail in Section~\ref{sec:6}, corresponds to the green curve. For this configuration, the BDDE is 22.68\% with HEVC as a reference. Thus, the energy demand is increased compared to HEVC and decreased compared to the reference configuration of VVC.

In Table~\ref{tab:HEVCVVC}, the evaluation of each metric for VVC is given for the coding configurations AI, LB, and RA in relation to HEVC. In general, the RD efficiency of VVC is increased significantly for each coding configuration and class. For AI, the BDR value is $-24.90\%$ on average. The bit rate is decreased at equal objective visual quality between $17.40\%$ for Class D and $38.76\%$ for Class F. However, the decrease of the bit rate is at the cost of a significantly increased computational complexity, which can be determined by the BDDE and BDDT values. On average, the decoding time is increased by $62.03\%$ and the decoding energy demand by $67.03\%$ for AI.

For LB, the bit rate is decreased by $31.01\%$, the decoding time is increased by $67.66\%$, and the decoding energy demand by $77.76\%$. For Class D, the decoding energy demand is increased by $134.35\%$, which is more than a doubling of the energy demand. 

For RA, the value of BDR is $-34.31\%$, of BDDT $71.05\%$, and BDDE $82.56\%$. In general, the increase of the decoding energy demand is disproportionately high compared to the decoding time demand for all measurements. This difference will be investigated in the following section. Furthermore, the complexity of RA is increased by a higher degree than of AI.

\section{Influence of SIMD Instructions on Decoding Time and Energy Demand}
\label{sec:4}

In this section, the influence of Single Instruction Multiple Data~(SIMD) instructions will be considered. The goal of SIMD instructions is to improve the performance when the same instruction is performed on multiple data elements~\cite{Chi2015}. Depending on the used CPU architecture, there are different SIMD instruction extensions available. In our measurements, the CPU uses AVX-2 SIMD instructions to decode the bit streams, if SIMD is enabled.

In order to not use SIMD instructions for the decoding process, we disabled a corresponding macro flag in the source code of VTM-7.0. The results of the measurements are shown in Table~\ref{tab:SIMD}, where the measurements of HM-16.20 are taken as a reference.

The results of Table~\ref{tab:SIMD} show that for AI, the energy demand is increased by $113.77\%$ and the time demand by $122.69\%$. For the LB configuration, the energy demand is increased by $158.84\%$ and the time demand by $162.73\%$. Finally, for the RA configuration, the time demand is increased by $211.76\%$ and the energy demand by $207.04\%$. Especially, for the bit streams of Class A1 and A2, which have a resolution of $3840\times2160$, the time and energy demand is increased the most. For Class A2 and RA configuration, the demand is increased by 249.26$\%$. Therefore, we can determine that the usage of SIMD improves the time and energy demand drastically. 

In general, the energy demand increases less than the time demand if SIMD instructions are disabled. Therefore, the usage of SIMD decreases the time demand by a higher degree than the energy demand, which is expected because additional hardware has to be used that contributes to the energy demand of the decoder. Disabling SIMD in VTM has the effect that execution time and energies can better be compared to HM with respect to the computational complexity, that means the number of operations. Hence, these results indicate that the overall computational complexity of VTM decoders is more than twice as high as for HM decoders.

\section{Influence of coding tools on the energy efficiency}
\label{sec:5}

In the following section, we will first describe some selected tools that are introduced in VVC. Afterwards, we will analyze the influence of these coding tools on the decoding energy and time demand.

The tool intra sub-partition~(ISP)~\cite{Luxan-Hernandez2019} is used for intra-predicted luma blocks and splits a block into multiple partitions in horizontal or vertical direction. The number of partitions (2 or 4) depends on the number of pels within the luma block and for each sub-partition, the same intra-prediction direction is used. 	
Matrix weighted intra prediction~(MIP) is an alternative intra prediction technique, which uses a matrix-vector multiplication for the intra prediction~\cite{Schaefer2019}. The tools MIP and ISP are evaluated on the AI configuration since these tools are based on intra-prediction. 

The following tools will be evaluated with the RA configuration.
Adaptive loop filter~(ALF) is a loop-filter that is used in VVC besides the deblocking filter and the sample adaptive offset (SAO) filter. For the filtering, two diamond shaped filters are used, which have the size $7 \times 7$ for the luma component and $5 \times 5$ for the chroma component~\cite{Tsai2013}. 

Adaptive motion vector resolution~(AMVR) improves the coding efficiency by using different levels of precision (e.g., quarter-pel, integer-pel, or 4-pel) for a motion vector~(MV) to code the motion vector difference~\cite{Liu2019}.
With bi-directional optical flow~(BDOF), the concept of optical flow is applied. With BDOF, it is possible to compensate motion within the block that regularly would be compensated by an additional splitting of the block~\cite{Alshin2010}.
Decoder-side motion vector refinement~(DMVR) increases the accuracy of MVs that are coded with the merge mode on the decoder side with bilateral matching~\cite{Gao2019}.
Low-frequency non-separable transform (LFNST) reduces the spatial redundancy for the prediction of the residuals. The usage of matrix multiplication reduces the redundancy of the transformation coefficients~\cite{Koo2019}.
In HEVC, DCT-II is used for coding the residuals. In VVC, other transformation schemes are used with the tool multiple transform set (MTS)~\cite{Nguyen2019}.
For inter-predicted CUs, it is possible to split a CU into two triangle-shaped blocks, which are split in the diagonal direction, with triangular partition mode~(TPM)~\cite{Blaeser2019}.

The analysis of the energy and time demand is evaluated in the following way. First, we disable each tool according to the recommendations of~\cite{JVET-Q0013}. Then, we calculate BDR and BDDE with respect to the reference configuration of VTM-7.0. Therefore, a positive value for BDR indicates that the usage of the tool increases the RD efficiency. For BDDT and BDDE, a positive value indicates that time and energy demand is decreased if the tool is used.

\begin{table}[!t]
\caption{List of all tools that are switched off to evaluate the compression performance. For the evaluation, YUV-BDR, BDDT, and BDDE are calculated in relation to the reference VTM-7.0 configuration. The tools are sorted in declining order of the BDDE value for the corresponding coding configuration. For further information on the tools, references are given.}
\label{tab:ToolOff}
\begin{center}
\vspace{-0.3cm}
\begin{tabular}{ c | c | c | c | c }
& \multicolumn{4}{c}{AI over VTM-7.0}\\
Tool &  YUV-BDR in \%  &  BDDT in \%  & BDDE in \%  & Ref.    \\
\hline \hline
 MIP   & $0.51$ & $~1.16~$ & $~0.80~$ & \cite{Schaefer2019}        \\
\hdashline
 ISP    & $0.54$ & $-1.34~$ & $-1.82~$ & \cite{Luxan-Hernandez2019} \\
\multicolumn{5}{c}{}\\
& \multicolumn{4}{c}{RA over VTM-7.0}\\
 Tool &  YUV-BDR in \% &  BDDT in \% & BDDE in \% & Ref.    \\
\hline \hline
 AMVR  & $1.29$ & $~2.84$ & $~2.28~$   & \cite{Liu2019}  \\
 TPM   & $0.39$ & $~2.32$ & $~2.02~$ & \cite{Blaeser2019}         \\
 LFNST & $0.73$ & $~2.09$ & $~1.63~$  & \cite{Koo2019}             \\
 MTS  & $0.61$ & $~1.69$ & $~1.38~$  & \cite{Nguyen2019}          \\
\hdashline
 BDOF   & $0.80$ & $-0.81$ & $-1.30~$  & \cite{Alshin2010}          \\
 DMVR  & $0.73$ & $-0.59$ & $-1.44~$  & \cite{Gao2019}            \\
 ALF   & $4.36$ & $-7.08$ & $-6.99~$  & \cite{Tsai2013}             \\
\end{tabular}
\end{center}
\vspace{-0.5cm}
\end{table}

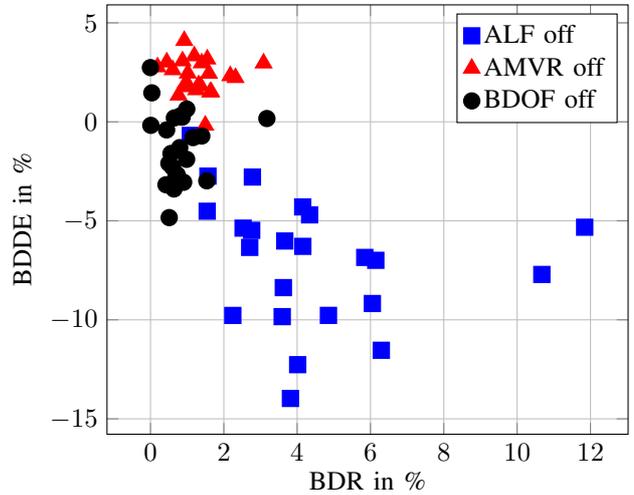
\begin{figure}[t!]
\begin{tikzpicture}
\begin{axis}[%
    xlabel={BDR in \%},
    ylabel={BDDE in \%},
	ymajorgrids,
	xmajorgrids,
   x label style={yshift=.1cm},
	legend cell align={left},
	scatter/classes={%
a={mark=square*,blue,mark size = 2.5pt,line width = 1.5pt},%
b={mark=triangle*,red,mark size = 2.5pt,line width= 1.5pt},%
c={mark=*,black,mark size = 2.5pt,line width = 1.5pt}}]
\addplot[scatter,only marks, %
  scatter src=explicit symbolic]%
table[meta=label]{
x y label
2.7592     -5.4795    a
2.5258     -5.3685    a
11.842     -5.3222    a
10.674     -7.7083    a
4.1542     -6.2886    a
3.6621     -6.0185    a
4.8536     -9.7751    a
1.5461      -4.515    a
2.7795     -2.7887    a
4.1517      -4.297    a
3.5928     -9.8385    a
6.1442     -6.9938    a
6.295     -11.537     a
4.0146     -12.264    a
3.6239     -8.3728    a
6.0507     -9.1758    a
5.8444     -6.8482    a
2.7061     -6.3421    a
1.5667      -2.741    a
2.2445     -9.7781    a
1.0872    -0.68485    a
4.3406     -4.6984    a
3.8178      -13.97    a
0.91741      1.7666   b
0.96235      1.9542   b
0.60633      2.6226   b
0.87376      3.0722   b
1.4939    -0.17918    b
1.6271      1.5023    b
2.3181      2.2276    b
1.3964      2.9702    b
0.44402      3.0134   b
1.0183       2.453    b
0.92071         4.1   b
1.594      2.4433    b
2.1753      2.3211   b
1.3686      1.8081   b
1.2454       1.612   b
1.3117      1.8631   b
0.76605      1.3427  b
1.5486      3.1636   b
1.1942      3.3244   b
1.6529      1.5124   b
0.19198      2.7687  b
1.0024      1.9222   b
3.084       2.962    b
0.6015     -2.2948     c
1.3972    -0.71682     c
3.1745     0.16659     c
0.98416     -1.8937    c
0.50114     -2.0971    c
0.55903     -1.5981    c
0.69825      -2.804    c
0.98833     0.64682    c
1.1603    -0.79812    c
0.6523     0.19785    c
0.0092673    -0.17761  c
0.86333     0.23405    c
1.5357     -2.9804     c
0.51046     -4.8363    c
0.69318     -2.7987    c
0.42245     -3.1771    c
0.90306     -3.0527    c
0.4394    -0.40956    c
0.8215     0.30784    c
0.63973     -3.3916   c
-0.0018377      2.7383 c
0.037544      1.4632  c 
0.72694     -2.6711   c  
0.79641     -1.3019   c
};
	\addlegendentry{ALF off}
	\addlegendentry{AMVR off}
	\addlegendentry{BDOF off}

\end{axis}
\end{tikzpicture}
\vspace{-0.3cm}
\caption{Evaluation of BDDE and BDR for each sequence of the CTC for RA configuration if the tools ALF, AMVR, and BDOF are disabled.}
\label{fig:seqwise}
\vspace{-.3cm}
\end{figure}

In Figure~\ref{fig:seqwise}, the BDDE and BDR values for each sequence are shown for the switched off tools ALF, AMVR, and BDOF in a scatter plot. It can be determined that ALF (blue squares) has a significant influence on the RD and energy efficiency. The values of BDDE are on average $-6.99\%$ and for BDR $4.36\%$. Therefore, the energy demand is increased if ALF is used by the decoder. Simultaneously, the bit rate is decreased by a lower degree than the energy demand. For BDOF (black points), we have a similar observation. The usage of the tool increases the energy demand and decreases the bit rate. For AMVR (red triangles), we see that the energy demand and the bit rate increase if the tool is switched off. Therefore, the usage of the tool improves the RD and the energy efficiency of VVC, which will be discussed later.

In Table~\ref{tab:ToolOff}, all presented tools are evaluated in terms of BDR, BDDE, and BDDT. For MIP, we can see that the decoding energy demand is increased by $0.80\%$ on average if the tool is switched off. Furthermore, it can be observed that the energy efficiency of VVC is improved by 5 tools and the RD efficiency by all tested tools. For the tools BDOF, DMVR, ISP, and ALF, the energy demand is decreased between $-7\%$ and $-1.30\%$, if the tool is switched off. If the tools MIP, AMVR, TPM, LFNST, and MTS are disabled, the energy demand is increased between $0.80\%$ and $2.28\%$. The time demand for these tools is increased between $1.16\%$ and $2.84\%$.

\begin{table*}[t!]

\caption{Evaluation of the proposed energy efficient coding configuration. The tools ALF, BDOF, DMVR, and ISP are disabled in the encoder. The configuration is compared to the reference RA configuration of HM-16.20 (left side) and VTM-7.0 (right side).}
\label{tab:energyefficient}
\vspace{-0.3cm}
\centering
\begin{tabular}{l | c |  c  | c || c | c | c}
     & \multicolumn{3}{c||}{RA over HM-16.20} & \multicolumn{3}{c}{RA over VTM-7.0} \\
 Class	     &  YUV-BDR in \% &  BDDT in \% & BDDE in \% &  YUV-BDR in \% &  BDDT in \% & BDDE in \%\\
	     \hline
	     \hline
A1 & $-33.96$ & $20.79$ & $28.20$   & $~6.06$ & $-24.19$ & $-24.28$ \\
A2 & $-34.05$ & $41.57$ & $51.10$   & $10.72$ & $-17.41$ & $-17.40$\\
B  & $-28.79$ & $27.84$ & $37.16$   & $~8.07$ & $-19.76$ & $-19.25$\\
C  & $-24.45$ & $44.59$ & $56.28$   & $~6.31$ & $-16.43$ & $-15.14$\\
D  & $-20.56$ & $71.52$ & $86.52$   & $~8.45$ & $-12.48$ & $-11.85$\\
F  & $-31.14$ & $37.37$ & $45.40$   & $19.50$ & $-16.35$ & $-16.58$\\
\hdashline
Mean & $-28.37$ & $40.88$ & $51.13$ & $~9.90$ & $-17.59$ & $-17.20$\\

\end{tabular}
\vspace{-0.3cm}
\end{table*}

Before performing the tool measurements, we expected that most tools increase the complexity and hence, the decoding energy. This is because each new tools adds complexity to the overall decoding process. However, as shown in Table~\ref{tab:ToolOff}, we can see that certain tools increase the energy efficiency. Both kinds of observations are discussed in the following. 

The changes in the energy demand for certain coding tools can be explained by the relative complexity. We use the bit stream analyzer proposed in \cite{Kraenzler2020} to analyze the reasons for the decreased or increased energy demand of several coding tools. For the tool ALF, the overall complexity and energy demand is increased due to additional filtering operations. An analysis of the tool AMVR reveals that the usage of fractional MVs is increased by approximately $17\%$ on average if AMVR is disabled. Consequently, the relative complexity of the decoder is reduced because less fractional pels have to be calculated by interpolation filter calculations. If DMVR is disabled, the energy demand is decreased by 1.44\%. The reason for this observation is that DMVR uses extensive calculations to further refine motion vectors on the decoder side. However, the refinement leads to higher complexity of the decoder, which leads to a higher energy demand.

For LFNST, we observed that the SAO filter is used significantly more frequently if LFNST is disabled. SAO is a loop filter that is used to reduce ringing artifacts \cite{Fu2012}. With bit stream feature analyzer software \cite{Kraenzler2020} it can be determined that for the luma component, the edge offset (EO) is used over 16\% more frequently in relation to the reference VTM encoding and for the chroma component, by over 25\%. Therefore, the relative complexity is decreased if LFNST is enabled because SAO is used less often. 

The increase of the energy demand for the tools BDOF and ISP can be explained by the additional computational complexity that is necessary to execute these tools. For MIP, the decreased energy demand is probably caused by the lower complexity compared to the conventional intra prediction modes. For MTS, we expected that the overall complexity barely changes because all transforms use the same number of calculations. In contrast, we observed energy savings, which can probably be attributed to a reduced number of residual coefficients and larger block sizes. For TPM, we even expect a slight increase in energy. To explain the observed savings, we suppose that the same reasons as for MTS hold.

As another observation, we find that the disabling of a specific tool has a similar effect on the decoding time and energy demand. The knowledge of the tool evaluation will be used in the following section to optimize the decoding energy demand of VTM decoded bit streams.

\section{Energy efficient coding configuration}
\label{sec:6}

As mentioned in previous sections, we propose a new coding configuration based on the results of the coding tool test. Therefore, we disable the tools ALF, BDOF, DMVR, and ISP for the RA coding configuration to decrease the energy demand of the decoder. In Table~\ref{tab:energyefficient}, the evaluated results of the proposed configuration are shown for the test set. On the left side of Table~\ref{tab:energyefficient} our proposed configuration is compared with HM-16.20 and on the right side with VTM-7.0. For the comparison with HEVC, it can be determined that the BDDE is on average 51.13\%. Therefore, the configuration has a significantly lower increase in energy demand compared to 82.56\% of Table~\ref{tab:HEVCVVC}. The BDR value of the proposed configuration is -28.37\%, which is less than in Table~\ref{tab:HEVCVVC}.

The comparison of the proposed coding configuration with the reference RA configuration reveals that with an additional bit rate of under 10\% it is possible to reduce the energy demand by over 17\%. For the sequences of Class A1, which have a 4K resolution, the energy demand is reduced by 24.28\%. Furthermore, the sequences of Class A2 and B show a higher BDDE value than the average. This indicates that the energy demand of sequences with at least Full-HD resolution decreases to a larger extent than low-resolution sequences. For the sequences of Class F, the BDR value is significantly higher with 19.50\% than the average value, which can be explained by the fact that these sequences show screen content.

The result of the proposed configuration shows that the energy demand can be reduced significantly while keeping the bit rate increase at a low degree. In the future, all coding tools of VVC can be analyzed, which will determine more tools that reduce the energy demand of the decoder if they are disabled. 

\section{Conclusion}
\label{sec:7}

In this paper, we analyzed the energy and time demand of VVC and determined that both increase significantly compared to HEVC. Furthermore, the energy demand is increased disproportionately high compared to the time demand. Afterwards, we analyzed the influence of SIMD instructions on the performance of the VTM decoder and revealed that the usage of SIMD decreases the time and energy demand significantly, where the time demand is decreased more than the energy demand. Finally, we evaluated some tools of the VVC standard and showed that MIP, AMVR, TPM, LFNST, and MTS improve the time and energy demand. The other tools that decrease the energy efficiency of the decoder are disabled in our proposed coding configuration. Thereby, we can reduce the energy demand of the VTM decoder by over 17\% on average. 

In future work, the evaluation of tools can be extended by including more coding tools of VVC. By that, it will be possible to determine more tools that can be used to reduce the energy demand of a coding configuration. Furthermore, the energy analysis can be evaluated with a real-time capable optimized implementation for VVC, which we expect to be available in future. 
Moreover, the decoding-energy-rate-distortion optimization that is proposed in~\cite{HerglotzHeindelKaup} for HEVC can be implemented in VVC for a further reduction of the energy demand.


\bibliographystyle{IEEEtran}


\end{document}